\documentclass[10pt]{article}

\usepackage{grffile}
\usepackage{ctable}

\usepackage{amssymb}
\usepackage{amsmath}
\usepackage{mathrsfs}
\usepackage[dvips]{epsfig}
\usepackage[small]{caption}
\usepackage{graphicx}
\usepackage[all]{xy}

\newtheorem{Lemma}{Lemma}[section]
\newtheorem{Theorem}{Theorem}

\newtheorem{Remark}[Lemma]{Remark}
\newtheorem{Definition}[Lemma]{Definition}
\newtheorem{Hypothesis}[Lemma]{Hypothesis}

\newenvironment{Proof}%
 {\begin{trivlist} \item[]{\bf Proof. }}%
 {\hspace*{\fill}$\rule{.4\baselineskip}{.4\baselineskip}$\end{trivlist}}

 {\begin{trivlist}\item[]\textbf{Acknowledgments.}}{\end{trivlist}}

\makeatletter\@addtoreset{figure}{section}\makeatother

\makeatletter \@addtoreset{equation}{section} \makeatother

\newcommand{\R}{\mathbb{R}}

\newcommand{\Z}{\mathbb{Z}}
\newcommand{\T}{\mathbb{T}}
\newcommand{\Q}{\mathbb{Q}}

\newcommand{\rmO}{\mathrm{O}}

\newcommand{\rmd}{\mathrm{d}}
\newcommand{\rme}{\mathrm{e}}

\makeatletter
\newsavebox{\@brx}
\newcommand{\llangle}[1][]{\savebox{\@brx}{\(\m@th{#1\langle}\)}%
  \mathopen{\copy\@brx\kern-0.5\wd\@brx\usebox{\@brx}}}
\newcommand{\rrangle}[1][]{\savebox{\@brx}{\(\m@th{#1\rangle}\)}%
  \mathclose{\copy\@brx\kern-0.5\wd\@brx\usebox{\@brx}}}
\makeatother

\definecolor{Green}{rgb}{0.,0.4,0.}

\renewcommand{\leq}{\leqslant}
\renewcommand{\geq}{\geqslant}

\newcommand{\Rmnum}[1]{\uppercase\expandafter{\romannumeral #1\relax}}

\def\Xint#1{\mathchoice
   {\XXint\displaystyle\textstyle{#1}}%
   {\XXint\textstyle\scriptstyle{#1}}%
   {\XXint\scriptstyle\scriptscriptstyle{#1}}%
   {\XXint\scriptscriptstyle\scriptscriptstyle{#1}}%
   \!\int}
\def\XXint#1#2#3{{\setbox0=\hbox{$#1{#2#3}{\int}$}
     \vcenter{\hbox{$#2#3$}}\kern-.5\wd0}}

\def\dashint{\Xint-}

\newfam\bifam
\font\tenbi=cmmib10 scaled \magstep1 \font\sevenbi=cmmib10 at 11pt
\font\fivebi=cmmib10 at 6pt \textfont\bifam = \tenbi
\scriptfont\bifam = \sevenbi \scriptscriptfont\bifam= \fivebi

\ifx\pdfoutput\undefined
   \pdffalse
\else
   \pdfoutput=1
   \pdftrue
\fi
\ifpdf
   \usepackage{graphicx}
   \usepackage{epstopdf}
\else
   \usepackage{graphicx}
\fi

\begin{document}

\begin{center}

{\fontsize{17}{17}\fontfamily{cmr}\fontseries{b}\selectfont{Depinning asymptotics in ergodic media}}\\[0.2in]
Arnd Scheel$\,^1$ and Sergey Tikhomirov$\,^2$\\
\textit{\footnotesize 
$\,^1$University of Minnesota, School of Mathematics,   206 Church St. S.E., Minneapolis, MN 55455, USA}\\
\textit{\footnotesize $\,^2$ St.Petersburg State University, 7/9 Universitetskaya nab., St. Petersburg,199034,  Russia}
\date{\small \today} 
\end{center}


\begin{abstract}
\noindent 
We study speeds of fronts in bistable, spatially inhomogeneous media at parameter regimes where speeds approach zero. We provide a set of conceptual assumptions under which we can prove power-law asymptotics for the speed, with exponent depending a local dimension of the ergodic measure near extremal values. We also show that our conceptual assumptions are satisfied in a context of weak inhomogeneity of the medium and almost balanced kinetics, and compare asymptotics with numerical simulations. 
\end{abstract}

\section{Fronts in inhomogeneous media --- a brief introduction and main results}\label{s:1}

We are interested in the speed of interfaces in spatially extended systems, separating stable or metastable states. A prototypical example is the Allen-Cahn or Nagumo equation, for the order parameter $u(t,x)\in\R$,
\begin{equation}\label{e:ac}
u_t=u_{xx}+(u-a)(1-u^2),\qquad x\in\R,\quad a\in(-1,1).
\end{equation}
This system possesses the spatially homogeneous, stable equilibria $u\equiv \pm 1$. Initial conditions with $u_0(x)\in(-1,1)$, $u_0(x)\to 1$, $x\to +\infty$, $u_0(x)\to -1$, $x\to -\infty$, converge to traveling waves for $t\to\infty$. In fact, \eqref{e:ac} possesses a unique (up to translation) traveling wave $u_*(x-st)$, connecting $-1$ and $1$, solving
\begin{equation}\label{e:actw}
u''+su'+(u-a)(1-u^2)=0,\qquad  u(x)\to \pm 1,\ x\to \pm\infty.
\end{equation}
The solution with $u(t=0,x)=:u_0(x)$ will then converge to a suitable translate of $u_*$ as $t\to\infty$,
\[
\sup_{x\in\R}|u(t,x)-u_*(x-st-\xi)|\to 0,\qquad t\to\infty,
\]
for some $\xi\in\R$.
In fact, this convergence is exponential in time, and the family of translates of $u_*$ can be viewed as a normally hyperbolic manifold in the phase space of, say, bounded uniformly continuous functions \cite{henry}. 

From this perspective, much of the information on a bistable medium is captured by a single number, the speed of propagation $s$, which is generally a function of $a$. In the specific example of the cubic, one first notices that for at $a=0$, often referred to as a ``balanced nonlinearity'', or the ``Maxwell point'', the speed vanishes. Somewhat surprisingly, the speed is in fact a linear function $s=\sqrt{2}a$ of the parameter, for this specific cubic nonlinearity. More generally, for bistable nonlinearities with three zeros,  $f(u_-)=f(u_+)=f(u_\mathrm{m})=0$, two stable zeros $f'(u_\pm)<0$ and one unstable zero $f'(u_\mathrm{m})>0$, one finds the existence of $u_*(x-st)$, with smooth dependence of speed and profile on parameters. In particular, one expects that, generically in one-parameter families of nonlinearities $f_a$, $s(a)=s_1(a-a_*)+\rmO\left((a-a_*)^2\right)$, where $a_*$ refers to the critical parameter 	value of a balanced nonlinearity. 

In fact, such speed asymptotics can be derived much more generally in systems of equations, provided that the linearization at a traveling wave $u_*(x)$ with speed $s=0$ for $a=a_*$ possesses an algebraically  simple eigenvalue $\lambda=0$; see the discussion in Section \ref{s:2} for more details. We therefore briefly write $s\sim \mu^1$, with $\mu=a-a_*$ the detuning parameter from criticality, thus  encoding the linear asymptotics near zero speed. In the sequel we refer to the parameter $a$ as \emph{imbalance}, the relation $s=s(a)$ as a \emph{speed-imbalance relation}. We normalize $a=0$ as the \emph{balanced} case with $s=0$, and write $a_\mathrm{c}=\inf_a\{s(a)>0\}$ for the \emph{critical} parameter. The generic situation described so far implies of course that $a_\mathrm{c}=0$. 

\paragraph{Pinning from inhomogeneity --- main questions.}

The above scenario changes qualitatively when inhomogeneities are present in the medium. Consider, for example, 
\begin{equation}\label{e:acin}
u_t=u_{xx}+(u-a)(1-u^2)+g(x,u;\varepsilon),\qquad x\in\R,\quad a\in(-1,1),
\end{equation}
with inhomogeneity, say, $g(x,\pm 1;\varepsilon)=0$ and $g(x,u;0)=0$. Examples of interest here are inhomogeneities that are periodic, quasi-periodic, or, more generally, ergodic with respect to $x$. In these situations, in particular for small $\varepsilon$, one then expects average speeds  $\bar{s}$ to exist, thus reducing essential properties of the medium again to a (average-)speed-imbalance relation $\bar{s}(a)$. Ergodicity of the medium here refers to the action of the shift on the medium, $g(\cdot,u;\varepsilon)\mapsto g(\cdot+\xi,u;\varepsilon)$ and encodes a transitivity and recurrence property (relative to an ergodic measure) for this ``dynamical systems'' of spatial translations, guaranteeing in particular the existence of averages for functions of the medium. 

\begin{figure}
\centering\includegraphics[width=0.9\textwidth]{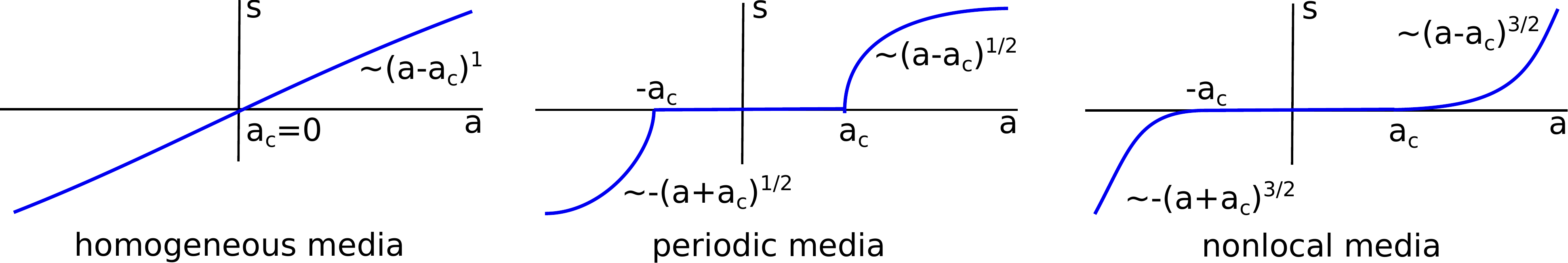}
\caption{Schematic illustration of speed versus imbalance parameter in the case of translation invariant, spatially periodic, and nonlocal media; compare \eqref{e:gamma}.}\label{f:1}
\end{figure}

The phenomenon of pinning refers to a situation when such an average speed vanishes for open sets of imbalances, that is,  $a_\mathrm{c}>0$ and $\bar{s}(a)=0$ for $a\in [0,a_\mathrm{c})$. One then refers to fronts at parameter values $a\in (0,a_\mathrm{c})$ as \emph{pinned fronts}, since small changes in the system will not allow the front to propagate; see Figure \ref{f:1} for a schematic illustration of speed-balance relations. 
Key questions concerning the phenomenon of pinning are:
\begin{enumerate}
\item when do we expect pinning, $a_\mathrm{c}>0$?
\item what is the size of the pinning region in terms of system parameters, $a_\mathrm{c}=a_\mathrm{c}(\varepsilon)$?
\item what are speeds near the pinning region, 
\begin{equation}
\bar{s}(a)\sim \mu^\gamma, \qquad \mu=a-a_\mathrm{c}, \quad \mbox{for some }\gamma>0?
\label{e:gamma}
\end{equation}
\end{enumerate}
The first item, (i),  refers to a rather general question and we give several examples where $a_\mathrm{c}>0$, that is, where a nontrivial pinning region occurs, at the end of this introduction. The second item (ii) refers to a dependence of $a_\mathrm{c}$ on system parameters. We are aware of only few cases where such dependencies are known analytically; see however our analysis in Section \ref{s:3} and the discussion, below. Our focus will be on (iii), striving to determine $\gamma$. 

\paragraph{Pinning from inhomogeneity --- outline of our main results.}

Our results here give a rather simple formula for $\gamma$ as a function of the \emph{effective dimension} of the medium near criticality. We give three different types of results:
\begin{enumerate}
\item an abstract skew-product formalism for depinning dynamics, deriving depinning asymptotics from Birkhoff's ergodic theorem;
\item a specific application to weakly inhomogeneous media;
\item some numerical results in agreement with the abstract results.
\end{enumerate}
From this point of view, the main open questions are how widely the abstract approach in (i) can be shown to be valid, beyond simple examples (ii) and numerics (iii). 
\begin{figure}
\centering\includegraphics[width=0.9\textwidth]{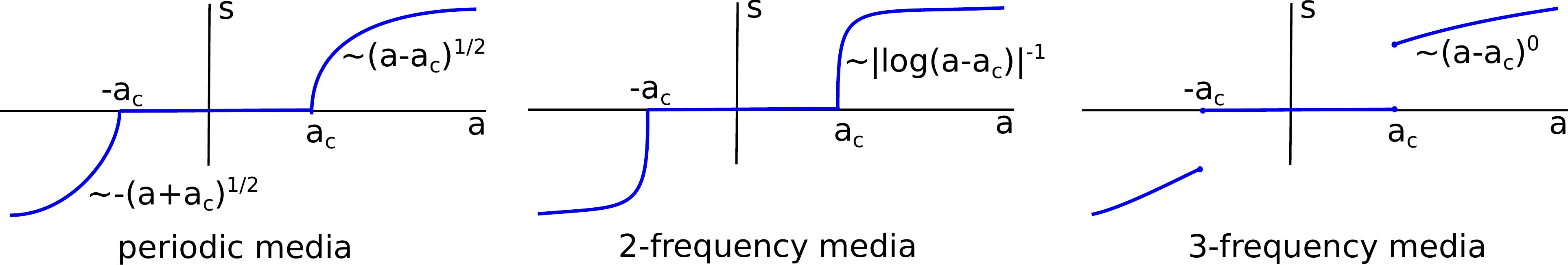}
\caption{Schematic illustration of speed versus imbalance parameter in the case of periodic and quasiperiodic media.}\label{f:1b}
\end{figure}

To state our main results briefly, consider an equation of the form \eqref{e:acin} with 
\[
g(x,u;\varepsilon)=h(\vartheta(x),u),
\]
where $\vartheta(x)=S_x(\theta)$ denotes a trajectory of a flow $S_x(\vartheta_0)$ on a smooth manifold $\mathcal{M}$ with ergodic measure $\nu$. The simplest example is a quasiperiodic medium
\begin{equation}\label{e:dim}
\vartheta(x)=\omega x \mod 1,\qquad \omega\in\R^m, \omega_j \mbox{ independent over } \R\setminus\Q,\qquad \vartheta(x)\in \mathcal{M}=\T^m=\R^m/\Z^m,
\end{equation}
such that $h(\cdot,u):\T^m\to\R$ is quasiperiodic with $m$ frequencies, and $\nu$ is simply Lebesgue measure on $\T^m$.

Our key assumption is that the dynamics near depinning can be reduced to a skew-product flow on $\mathcal{M}\times\R$, with flow $S_x$ acting on $\mathcal{M}$ as determined by the medium, and reduced flow $\xi'=s(S_\xi(\theta);\mu)$, $s:\mathcal{M}\times\R\to\R$.

Let us assume that $a=a_\mathrm{c}$, $\mu=0$ as in \eqref{e:gamma} is critical. More precisely, we assume that at $\mu=0$, ${s}(\vartheta_*;0)=0$ and  ${s}(\vartheta;0)>0$ for $\vartheta\neq \vartheta_*$ in $\mathcal{M}$. Moreover, we assume nondegenerate criticality, 
\[
D^2_\vartheta s(\vartheta_*;0)>0,\qquad  \partial_\mu s(\vartheta_*;0)>0.
\]
Then we find, for $\nu$-almost all media,
\begin{equation}\label{e:g}
\bar{s}(\mu)\sim \mu^{\gamma}, \qquad \gamma = \max\{1-\frac{\kappa}{2},0\}, \quad \kappa\neq 2,
\end{equation}
where $\kappa$ is the dimension of the ergodic measure $\nu$ at $\vartheta_*$. Here, $\gamma=0$ refers to discontinuous behavior at $\mu=0$, that is, $\lim_{\mu\searrow 0} \bar{s}(\mu)>0$. For $\kappa=2$, we find logarithmic asymptotics,
\begin{equation}\label{e:gc}
\bar{s}(\mu)\sim \frac{1}{|\log(\mu)|}, \qquad \kappa=  2.
\end{equation}
For quasiperiodic media, $\kappa$ simply stands for the number of frequencies, and our results predict hard depinning, that is,  discontinuous speeds, for three or more frequencies; see Figure \ref{f:1b}. 

To our knowledge, such asymptotics for speeds are new beyond periodic media. 
Existence and propagation of fronts has been established in various contexts; see  \cite{nolen2} for existence of speed in random and ergodic media, with ignition type nonlinearities, not allowing for pinning, and see \cite{matano,nopinning} for the bistable case with no-pinning assumption, and \cite{xin0} for an overview.
For depinning asymptotics, we refer to \cite{stat2,kardar,stat3,stat4,vd} for results using renormalization group theory for random media, and for results in specific model problems. 

In the remainder of this introduction, in order to give some context to the somewhat general setup here, we review several special cases that are  understood to some degree.

\begin{figure}
\centering
\includegraphics[width=0.85\textwidth]{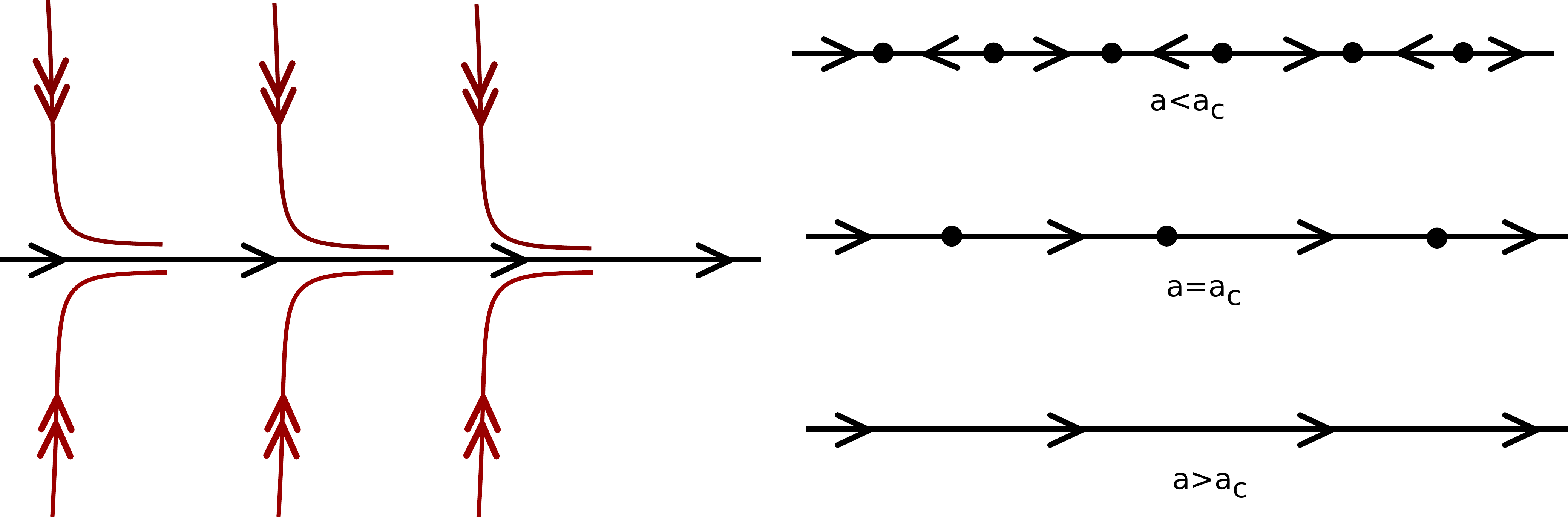}
\caption{Schematic illustration dynamics in phase (function) space. Homogeneous media with a normally attracting (red) invariant manifold (black) of translated traveling wave profiles (left); the invariant manifold with spatially periodic vector field for periodic media, showing the depinning transition as a saddle-node (right).}\label{f:3}
\end{figure}

\paragraph{Depinning in periodic media.}  The key ingredients to our main result is the description of front dynamics through positional dynamics, and the presence of a saddle-node bifurcation. We illustrate those ingredients in the well understood case of periodic media. Similar to the translation invariant case, one may expect, for instance when $\varepsilon\ll 1$ in \eqref{e:acin},  that there exists a family of interfaces parameterized by a position variable $\xi\in\R$, which form a normally hyperbolic manifold in a suitable function space such as the bounded uniformly continuous functions.  In the pinning regime, this manifold contains pinned fronts as equilibria, and heteroclinic orbits between those equilibria; see Figure \ref{f:3} for a schematic picture and \cite{hamel}  for general results towards establishing existence of such manifolds in a non-perturbative setting. 
Parameterizing the manifold by $\xi\in\R$, one then infers positional dynamics 
\[
\xi'=s(\xi;a),\quad s(\xi+P;a)=s(\xi;a),
\]
with pinning region given by the values of $a$ where $s$ possesses a zero. Generically, zeros will disappear in a saddle-node bifurcation, with local expansion 
\[
\xi'=\alpha_1 (a-a_\mathrm{c})+\alpha_2 (\xi-\xi_0)^2+\ldots
\]
One readily finds that for $\alpha_1,\alpha_2>0$, say, and $a\gtrsim a_\mathrm{c}$, the passage time $T$ from $\xi_0-\delta$ to $\xi_0+\delta$, $\delta>0$, fixed, scales as $T\sim (a-a_\mathrm{c})^{-1/2}$, such that the average speed scales 
\begin{equation}\label{s:sn}
\bar{s}(a)\sim (a-a_\mathrm{c})^{1/2},
\end{equation}
that is, $\gamma=1/2$ in (iii), or $\kappa=1$ in \eqref{e:g}, consistent with the one-dimensional nature of a periodic medium, $m=1$ in \eqref{e:dim}. Results that establish asymptotics of this type are abundant in the literature, and we mention here \cite{dill,lw} for rigorous results with weak inhomogeneities and \cite{physlatt} for explicit asymptotics in lattices.

\paragraph{Slowly varying media --- intuition on pinning.}

The most intuitively accessible scenario are slowly varying media, for instance
\[
u_t=u_{xx}+(u-a)(1-u^2) + g(x,u;\varepsilon), \qquad g(x,u;\varepsilon)= A(\varepsilon x)(1-u^2),
\]
where $A(y)\in(-1,1)$ and $\varepsilon \ll 1$. Intuitively, since the front interface experiences an almost constant medium with effective imbalance $a_\mathrm{eff}(\xi)=a-A(\varepsilon \xi)$, one expects that speeds depend on the front position $\xi$ through $s=s(a-A(\varepsilon\xi))$, where $s(\cdot)$ is the speed-balance relation from the spatially homogeneous case $\varepsilon=0$. One then infers a leading-order differential equation for the front position 
\[
\xi'=s(a-A(\varepsilon\xi)),
\]
which exhibits equilibria whenever $a-A(\zeta)$ has at least one zero. Phenomenologically, front propagation is blocked at locations where $s(a-A(\varepsilon x))=0$. Suitably defined averages of the speed can now vanish for open sets of the parameter $a$ since zeros  can be robust. 
Although this result appears intuitive, and although a formal expansion in $\varepsilon$ gives such a result to leading order, we are not aware of a result that rigorously establishes such a description; see however \cite{yip1,yip2} for results on depinning bifurcations in this context.

\paragraph{Rapidly varying media, homogenization, and exponential asymptotics.}
In the opposite direction, one can consider rapidly varying media $  g(x,u;\varepsilon)= A(x/\varepsilon )(1-u^2)$. When variations of the medium are fast compared to the scale of variations for the front, one would hope to replace $A$ by its (local) average, obtaining a homogenized equation with recovered translational invariance. Therefore, the pinning region is trivial in the averaged equation, but one expects non-trivial pinning regions for $\varepsilon>0$. In fact, assuming $A$ periodic, with $\dashint A=0$, say,  one can see that $a_\mathrm{c}(\varepsilon)>0$, generically, in the class of smooth periodic functions, following the steps in \cite{fisc}. For analytic functions $A(\xi)$, the estimates there imply $a_\mathrm{c}=\rmO(\rme^{-c/\varepsilon})$ for some $c>0$, implying an extremely small pinning region. Similar considerations also apply to the case where inhomogeneitiy is reflected in a space-dependent diffusivity. 
Still following the ideas in \cite{fisc}, we can also think of a spatial finite-differences discretization of \eqref{e:ac} as encoding a spatially periodic dependence of the diffusion coefficient; see also \cite{vVs}. Fine discretizations would then lead to extremely small pinning regions in the above sense.

\paragraph{Lattice dynamical systems.}

Pinning is also present in lattice differential equations,
\begin{equation}\label{e:lds}
u_{j,t}=d(u_{j+1}-2u_j+u_{j-1})+f_a(u_j),\qquad j\in\Z,
\end{equation}
where more general, say next-nearest neighbor coupling is also possible. Here, $f_a(u)$ is a general bistable nonlinearity, for instance the function $f_a(u)=(u-a)(1-u^2)$ from above. Pinning regions are explicit in the case of lattice differential equations with piecewise linear nonlinearities \cite{keanlattice}, but the general phenomenon had been noticed much earlier in the literature; see for instance  \cite{xin,xin2,zinner} for spatially periodic media and \cite{cb,mp2,mp} for results on lattice dynamical systems.

Lattices can be thought of as inherently spatially inhomogeneous, periodic media, where the translation symmetry is reduced to the discrete group $\Z$. They can be rigorously approximately embedded into spatially periodic media of the form
\begin{equation}\label{e:acd}
u_t=(d_\mathrm{p}(x)u_x)_x +f_a(u),\qquad d_\mathrm{p}(x)=d_\mathrm{p}(x+1),
\end{equation}
see \cite{vVs}.

\begin{figure}
\includegraphics[width=0.95\textwidth]{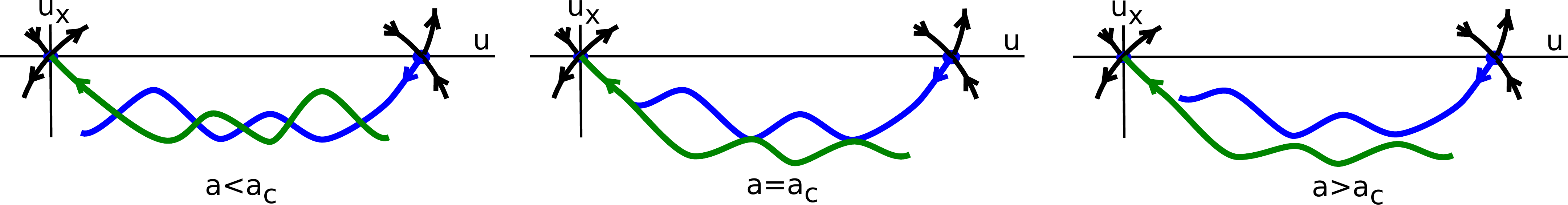}
\caption{Schematics of phase portraits for period maps associated with  \eqref{e:acd} or \eqref{e:lds}; depinning transition corresponds to heteroclinic tangency.  }\label{f:2}
\end{figure}

Pinning regions can be associated with regions where stationary solutions exist. Such stationary solutions solve a two-term recursion 
$d(u_{j+1}-2u_j+u_{j-1})+f_a(u_j)=0$, or a  time-periodic differential equation, $(d_\mathrm{p}(x)u')'+f_a(u)=0$ in the case of \eqref{e:acd}. Both define a diffeomorphism of the plane, the latter after passing to the time-one map, the first by simply writing the recursion as a first-order recursion in the plane. Both possess heteroclinic orbits between hyperbolic equilibria at $a=0$. Since these heteroclinic orbits are generically transverse, they occur for open intervals of the parameter $a$, thus implying pinning; see Figure \ref{f:2} for a schematic picture of heteroclinic orbits unfolding when varying $a$. We refer however to \cite{keanlattice,pelsan} for examples of discrete systems which do not exhibit pinning. 

\paragraph{Nonlocal coupling.}
We also mention a curious phenomenon that arises when interpreting \eqref{e:lds} as an equation on the real line, 
\begin{equation}\label{e:nll}
u_t(t,x)=d(u(t,x+1)+u(t,x-1)-2u(t,x))+f(u(t,x))=2d (-u+\mathcal{K}*u)(t,x) + f(u(t,x)),
\end{equation}
with $\mathcal{K}=(\delta_{-1}+\delta_{+1})/2$, Dirac-$\delta$ functions shifted by $\pm 1$. Naturally \eqref{e:nll} decouples into a family of equations on $x_0+\Z$, each equivalent to \eqref{e:acd}, for which we expect non-trivial pinning and depinning with depinning exponent asymptotics $\gamma=1/2$ in \eqref{e:gamma}. Curiously, this happens to be a very degenerate situation, as far as asymptotics are concerned, as demonstrated in \cite{fs}. For smooth convolution kernels $\mathcal{K}$, the results there demonstrate that one has $\gamma=3/2$ for some simple kernels with rational Fourier transform, which formally corresponds to an ergodic dimension $\kappa=-1$ in \eqref{e:g}. Numerical results strongly suggest that $\gamma=3/2$ for all smooth enough kernels, and $\gamma>3/2$ for kernels with strong singularities at the origin; see Figure \ref{f:1} for a schematic comparison.

\paragraph{Outline.} We give a precise statement and prove our main result in Section \ref{s:2}. Section \ref{s:3} contains an example which allows for a verification of our main assumptions in the case of weak inhomogeneities. We study more general situations numerically in Section \ref{s:4} and conclude with a brief discussion, Section \ref{s:5}.

\paragraph{Acknowledgements.} A. Scheel was partially supported through NSF grants DMS-1612441 
and DMS-1311740, through a DAAD  Faculty Research Visit Grant, WWU Fellowship, 
and a Humboldt Research Award. S. Tikhomirov was supported by Saint-Petersburg State University research grant 6.38.223.2014, and RFBR 15-01-03797a and by "Native towns", a social investment program of PJSC "Gazprom Neft".

\section{Depinning --- abstract result}\label{s:2}

We consider an abstract system, 
\begin{align}
U_t&=F(U,\theta;\mu),\nonumber\\
\theta_t&=0,\label{e:U}
\end{align}
where $U\in X$, a Banach space, is the state vector,  $\theta\in\mathcal{M}$, a smooth compact manifold, encodes the medium, and $\mu\in\R$ encodes the depinning parameter. The equation for $U$ is understood to generate a smooth semiflow, although this is not a technically relevant assumption in addition to the hypotheses listed below, rather relevant for their verification in specific examples.  We assume that the system possesses a translation symmetry  acting on $U$ and $\theta$, 
\[
\mathcal{T}_\zeta=\mathrm{diag}\,(T_\zeta,S_\zeta), \qquad \zeta\in\R,
\]
with action $S_\zeta$ being a smooth flow on $\mathcal{M}$, and $T_\zeta$ encoding translations of profiles.

\begin{Hypothesis}[invariant manifold]\label{h:1}
We assume that there exists a family of smooth manifolds $\mathcal{N}_\mu\subset X\times \mathcal{M}$ diffeomorphic to $\R\times\mathcal{M}$, invariant under the action of $\mathcal{T}_\zeta$. The manifold $\mathcal{N}_\mu$ carries a flow such that all trajectories are solutions to \eqref{e:U}. In the ``coordinates'' $(\xi,\theta)\in\R\times\mathcal{M}$, the translations act as $(\xi,\theta)\mapsto (\xi+\zeta,S_\zeta(\theta))$, and the flow on $\mathcal{N}_\mu$ is generated by a $C^2$-vector field 
\begin{equation}\label{e:skew}
\xi'=s(S_\xi(\theta);\mu),\qquad \theta'=0.
\end{equation}
\end{Hypothesis}

Typically the existence of such an invariant manifold with smooth flow, smoothly depending on parameters, will be obtained by establishing normal hyperbolicity. We will give an example in the next section. The $\xi$-direction is associated with the translation group, parameterizes translates in $U$-space. As a consequence, $s$ is naturally interpreted as a speed. Zeros of $s$ are pinned profiles, $s>0$ on $\mathcal{M}$ corresponds to a depinned situation.

Our next hypothesis is concerned with the flow on the reduced manifold.

\begin{Hypothesis}[critical, generic pinning, and depinning]\label{h:2}
There is a unique $\theta_*\in\mathcal{M}$ such that
\begin{align*}
s(\theta;0)&>0 \mbox{ for }\theta\neq \theta_*,\\
s(\theta_*;0)&=0,\\
\partial_\mu s(\theta_*;0)&>0,\\
D^2_\theta s(\theta_*;0)&>0,
\end{align*}
that is, in words, we have positive drift speed $s$ except at a non-degenerate minimum $\theta_*$, whose value increases linearly with $\mu$. 
\end{Hypothesis}

Our last assumption concerns the medium. 

\begin{Hypothesis}[ergodic inhomogeneities and dimension]\label{h:3}
We assume that the flow $S_\zeta$ is ergodic with respect to an invariant measure $\nu$ on $\mathcal{M}$ and that the local dimension $\kappa$  of $\nu$  at $\theta_*$, as defined below, exists.
\end{Hypothesis}

\begin{Definition}[local dimension]
We say that the measure $\nu$ at a point $y_*$ has the dimension $\kappa\geq 0$, if there are constants $c,C>0$ such that the measure of balls of radius $r$ can be estimated through
\[
cr^\kappa\leq \nu(B_r(y_*))\leq Cr^\kappa.
\]
\end{Definition}
We note that the definition of dimension here is more restrictive than the more common definition $\kappa=\lim_{r\to 0} \log \nu(B_r)/\log r$. In other words, the local dimension might not exist for many ergodic measures. We suspect that results of the type derived here are possible for weaker characterizations of a local dimension, allowing for instance for slowly varying constants $c=c(\log r), C=C(\log r )$ in our characterization. We note that the definition of dimension used here is independent of Lipshitz coordinate changes.
With these three assumptions, we are now ready to state a precise version of our main abstract result. 

\begin{Theorem}\label{t:1}
Assume Hypotheses \ref{h:1}--\ref{h:3} on invariant manifolds, pinning, and ergodicity of the medium, respectively.  Then we have, for $\nu$-almost every medium $\theta\in\mathcal{M}$, and $|\mu|$ sufficiently small,
\begin{itemize}
\item \emph{pinning} for $\mu<0$, that is,  $\xi(t)$ is bounded for $t\in\R$;
\item \emph{depinning} for $\mu>0$, that is, $\xi(t)\to \pm\infty$ for $t\to\pm\infty$;
\item \emph{depinning asymptotics} depending on the local dimension, that is, $\lim_{T\to\infty}\xi(T)/T=\lim_{T\to-\infty}\xi(T)/T=\bar{s}(\mu)$ exist, with 
\begin{equation}\label{e:sasy}
\displaystyle{\bar{s}(\mu)\sim\left\{
\begin{array}{ll}
\mu^{1-\kappa/2},& \kappa<2,\\
(|\log(\mu)|^{-1},& \kappa=2,\\
1,& \kappa>2,\\
\end{array}\right.}
\end{equation}
as $\mu\to 0$, where the similarity sign refers to inequalities bounding the left-hand side in terms of the right-hand side from above and below with  $\mu$-independent nonzero constants.
\end{itemize} 
\end{Theorem}

\begin{Proof}
To prove depinning, note that $s>0$ for $\mu>0$, small. Hence, by compactness of $\mathcal{M}$, $s>s_\mathrm{min}$ and $\xi'>s_\mathrm{min}$, which proves the claim. To prove pinning, note that $s<0$ in an open neighborhood of $\theta_*$ when $\mu<0$, which, according to Hypothesis \ref{h:3} has positive but not full measure. Therefore, for $\nu$-almost every medium, $s(S_{\xi(t)}(\theta);\mu)<0$ and  $s(S_{\xi(t')}(\theta);\mu)>0$  for arbitrarily large values of $t,t'>0$ or $t,t'<0$, implying that $s(S_{\xi(t)}(\theta);\mu)$ changes sign. As a consequence, the trajectory $\xi(t)$ converges to an equilibrium and remains bounded. It remains to establish depinning asymptotics. We therefore solve the equation for $\xi$ explicitly, using that $s>0$,
\[
\int_0^\xi \frac{1}{s(S_\zeta(\theta);\mu)} \rmd \zeta = T.
\]
Therefore, 
\[
\bar{s}(\mu)=\left(\lim_{\xi \to \infty}\frac{1}{\xi}\int_0^\xi \frac{1}{s(S_\zeta(\theta);\mu)}\rmd \zeta	\right)^{-1},
\]
whenever the limit exists. Inspecting this integral, we notice that Birkhoff's ergodic theorem guarantees that the ``temporal'' $\zeta$-average exists $\nu$-almost everywhere and can be replaced by the space average over $\mathcal{M}$ weighted with the ergodic measure $\nu$, for almost all media $\theta$ (alias initial conditions for the spatial flow $S_\zeta$),
\[
\bar{s}=\left(\int_{\mathcal{M}} \frac{1}{s(\vartheta;\mu)}\rmd\nu(\vartheta)\right)^{-1}.
\]
One realizes that the integral gives a bounded contribution in a complement of a fixed small ball  $B_\delta(\theta_*)$ of radius $\delta>0$,  centered at the singularity $\vartheta_*$ of $s$, as $\mu\to 0$. We therefore find the leading-order contribution
\[
\bar{s}\sim\left(\int_{B_\delta(\theta_*)} \frac{1}{s(\vartheta;\mu)}\rmd{\nu}\right)^{-1}
.\]
Further simplifying the calculation, we can choose coordinates according to the Morse Lemma $\tilde{\vartheta}=\Psi(\vartheta;\mu)$, such that $\theta_*=0$ and $s(\vartheta;\mu)=\mu+|\vartheta|^2$ in a small neighborhood of the origin, possibly also reparameterizing the parameter $\mu$. Smoothness of the coordinate change ensures that the dimension of the transformed measure is unchanged. This leads to the integral asymptotics
\[
\bar{s}\sim \left(\int_{B_{\delta'}(0)} \frac{1}{\mu+|\vartheta|^2}\rmd\tilde{\nu}\right)^{-1}
,\]
with transformed measure $\tilde{\nu}$. Here, we also changed the domain of integration to a small ball $B_{\delta'}(0)\subset \psi(B_\delta(\theta_*);\mu)$ centered at the origin, which again does not affect asymptotics since contributions outside of a small neighborhood are bounded. Rescaling, we may assume $\delta'=1$. 

In the case of $\kappa$-dimensional Lebesgue measure, the right-hand side can now be evaluated explicitly to find the result. Alternatively, one would obtain the asympmtotics by scaling $\theta=\sqrt{\mu}\tilde{\theta}$ and exploiting scaling properties of Lebesgue measure. This latter approach can be exploited in our context of measures with possibly fractal dimension.

In order to estimate the integral, define $D_\ell=B_{\rho^{\ell}}(0)\setminus B_{\rho^{\ell+1}}(0)$, for $\rho<1$ sufficiently small, and find 
\[
\tilde{c}\rho^{\kappa \ell}\leq \sum_{\ell=1}^\infty\int_{D_\ell}\rmd\nu \leq \tilde{C}\rho^{\kappa \ell},
\]
where $\tilde{c}=c-\rho^\kappa C>0$, $\tilde{C}=C-\rho^\kappa c$ in terms of $c,C$ from Hypothesis \ref{h:3}.

Suppose now that $\kappa<2$. We evaluate the integral by first decomposing into sums, $\int_{B_1(0)}=\sum_{\ell=0}^\infty \int_{D_\ell}$, which gives
\[
\int_{B_\delta(\vartheta_*)} \frac{1}{s(\vartheta;\mu)}\rmd\nu 
\sim \sum_{\ell=0}^\infty \int_{D_\ell} \frac{1}{\mu+|\vartheta|^2}\rmd\tilde{\nu}.
\]
In a region $D_\ell$, the integrand can be estimated from above and below with $\mu$-uniform constants as $(\mu+\rho^{2\ell})^{-1}$, which gives
\[
\int_{B_\delta(\vartheta_*)} \frac{1}{s(\vartheta;\mu)}\rmd\nu 
\sim 
\sum_{\ell=0}^\infty \rho^{\kappa \ell}\frac{1}{\mu+\rho^{2\ell}}\sim \int_{x=1}^\infty \rho^{\kappa x}\frac{1}{\mu+\rho^{2x}}\rmd x\sim \int_0^1r^{\kappa-1}\frac{1}{\mu+r^2}\rmd r \sim \mu^{\kappa/2-1}.
\]
Here we used the integral criterion for sums to reduce the sum to an elementary integral, which can be computed after the substitution  $r=\rho^{x}$. 

For $\kappa>2$, the resulting integral is uniformly bounded away from zero,
\[
\int_{B_\delta(\vartheta_*)} \frac{1}{s(\vartheta;\mu)}\rmd\nu 
\sim \int_{x=1}^\infty \rho^{\kappa x}\frac{1}{\mu+\rho^{2x}}\rmd x\sim 1,
\]
and for $\kappa=2$ we find logarithmic asymptotics,
\[
\int_{B_\delta(\vartheta_*)} \frac{1}{s(\vartheta;\mu)}\rmd\nu 
\sim \int_{x=1}^\infty \rho^{2 x}\frac{1}{\mu+\rho^{2x}}\rmd x\sim |\log(\mu)|.
\]
\end{Proof}
Intuitively, for larger dimensions, regions where the front is almost pinned are less frequently explored, such that the front encounters regions where it is very slow less frequently.
In this sense, our result can be thought of as simply describing the effect of extreme-value statistics on the average speed of the front: the front will experience an effective slow down near the depinning threshold only when the extremely small values of the speed are explored sufficiently frequently in a power law scaling sense. Depinning is ``soft'', with small speeds near the threshold, for $\kappa<2$, and ``hard'' for $\kappa>2$, with speeds $\rmO(1)$ immediately after depinning.

The simplest examples are of course quasi-periodic media, where $\mathcal{M}=\R^\kappa/\Z^\kappa$, the $\kappa$-dimensional torus, with irrational flow preserving $\kappa$-dimensional Lebesgue measure. Depinning occurs with exponent $1/2$ for one frequency, with logarithmic asymptotics for two frequencies, and we find hard depinning for more than two frequencies. The degenerate case of $\kappa=0$ comprises the case of a Dirac measure at $\theta_*$, which corresponds to a translation-invariant medium, where we expect smooth asymptotics for the speed $s\sim\mu$, consistent with our expansions for $\kappa=0$.  

\begin{Remark}[autonomous formulation]\label{r:geo}
The reduced equation can be written in  somewhat more compact form. Introducing the shifted medium $S_\xi(\theta)=:\psi\in\mathcal{M}$ as a new variable, we find the system
\[
 \dot{\xi}=s(\psi), \qquad \qquad \dot{\psi} = s(\psi)\sigma(\psi),
\]
where $\sigma$ is the vector field associated with the flow $S$. Geometrically, the flow for $\psi$ is the same flow as the ergodic flow of the medium, scaled by the scalar local velocity associated with the medium. Pinning occurs when the flow for $\psi$ possesses hypersurfaces of equilibria. Depinning occurs when those equilibria disappear, generically in the form of small shrinking ellipsoid. Slow speeds are caused by long passage times of trajectories in those regions. 
\end{Remark}

\begin{Remark}[Birkhoff-a.e.] 
In the theorem, we exploit the measure in order to use Birkhoff's ergodic theorem. While Birkhoff's theorem guarantees ergodic averages to converge to the average over the ergodic measure, convergence often holds for more trajectories, for instance trajectories with positive Lebesgue measure; see for instance \cite{ce}. In the simple example of quasiperiodic media with irrational flow on a torus, Birkhoff's theorem holds of course for all initial conditions, such that results are valid for all rather than almost every medium.
\end{Remark}

In the next section, we address how the assumptions made here can be verified in a prototypical example of weakly inhomogeneous media.


\section{Depinning with weak inhomogeneities --- an example}\label{s:3}

Our goal here is to provide an example where the hypotheses of Theorem \ref{t:1} can be verified. Consider therefore the classical bistable Nagumo equation
\begin{equation}\label{e:n}
u_t=u_{xx}+(u+\mu)(1-u^2)+\varepsilon \alpha(x;\theta)g(u).
\end{equation}
Note that for $\mu=\varepsilon=0$, the equation possesses a family of standing fronts, given explicitly as translates of  $u_*(x)=\tanh(x/\sqrt{2})$. 

It is natural to assume that the weak inhomogeneity $\varepsilon \alpha(x;\theta)$ depends on the variable $\theta\in\mathcal{M}$ in a smooth fashion, on compact intervals of $x$. Thinking for instance about solving \eqref{e:n} in spaces of bounded, uniformly continuous functions $BC^0_\mathrm{unif}(\R)$, equipped with the supremum norm, one would then like to assume even stronger \emph{uniform} smooth dependence, that is, smoothness of the map $\theta\mapsto \alpha(\cdot;\theta)\in BC^0_\mathrm{unif}(\R)$, say. This, however, would conflict with allowing for spatially ``chaotic'' media, which would incoporate some sensitive dependence of the spatial ``trajectory'' $S_{x}(\theta)$ on $\theta$, typically through exponential growth in the linearization measured by Lyaupunov exponents. As a consequence, there will exist locations $x_j$ such that $d(S_{x_j}(\theta_j),S_{x_j}(\theta_\infty))\geq \delta>0$ when $\theta_j\to \theta_\infty$, such that the map $\theta\mapsto \alpha(\cdot;\theta)\in BC^0_\mathrm{unif}(\R)$ cannot be continuous (it would of course be continuous in a local topology).  In order to recover smooth dependence of solutions on $\theta$, we exploit the fact that the bounded function $\alpha$ is multiplied by a term $g(u)$. Making suitable assumptions on $g$ will ensure exponential decay,  that is, the effect of the medium vanishes at the asymptotic states of the front. 

The following hypothesis quantifies this divergence. Therefore, define the function space of exponentially growing continuous functions $BC^0_\eta$ as the image of $BC^0_\mathrm{unif}$ under the multiplication map $u(\cdot)\mapsto \rme^{\eta|\cdot|}u(\cdot)$, that is, we allow for exponential growth with rate $\eta>0$.

\begin{Hypothesis}[small Lyapunov exponent inhomogeneity]
We assume that the function $\alpha:\R\times \mathcal{M}\to\R$ is smooth, bounded, and equivariant in the sense that $\alpha(x;\theta)=\alpha(0;S_{x}(\theta))$ for a smooth flow $S$ on $\mathcal{M}$ with ergodic invariant measure $\nu$.  Moreover, we assume that the map
\[
I_\alpha:\mathcal{M}\to  BC^0_\eta,\qquad \theta\mapsto \alpha(0;S_{\cdot}(\theta)),
\]
is of class $C^2$ for some $\eta\in\R$. \label{h:l0}
\end{Hypothesis}

We next state our assumption on the nonlinearity in the perturbation $g$.

\begin{Hypothesis}[vanishing tail corrections]\label{h:11}
We assume that the effect of inhomogeneities vanishes to zeroth and first order at the asymptotic states, $g(\pm 1)=g'(\pm 1)=0$, $g\in C^2$.
\end{Hypothesis}

\begin{Theorem}\label{t:2}
Consider equation \eqref{e:n} for $\varepsilon$ sufficiently small, with $\alpha$ satisfying Hypothesis \ref{h:l0} and $g$ satisfying Hypothesis \ref{h:11}. Furthermore, assume \emph{tail-Lyapunov dominance}, $\delta\geq \eta$, where $\delta=\sqrt{2}$ is the decay rate of the front and $\eta$ was specified in Hypothesis \ref{h:l0}. 

Then Hypothesis \ref{h:1} is satisfied, that is, there is a flow-invariant normally hyperbolic invariant manifold with an equivariant flow as stated there. Moreover, the reduced vector field $s:\mathcal{M}\times\R\to \R$ is smooth in $\varepsilon,\mu$ and $C^2$ in $\theta$, and possesses the expansion
\[
s(\theta;\varepsilon,\mu)=s_\varepsilon(\theta) \varepsilon + s_\mu \mu + \rmO\left(\varepsilon^2+\mu^2\right),
\]
where 
\begin{align*}
s_\varepsilon(\theta) &= \frac{ \left\langle u_*'(y),\alpha(y;\theta)g(u_*)\right\rangle}{\langle u_*',u_*'\rangle},\\
s_\mu & = \frac{\left\langle u_*'(y),g(u_*)\right\rangle}{\langle u_*',u_*'\rangle}.
\end{align*}

%
%
%
%
%
%
%
%
\end{Theorem}
\begin{Proof}
In our ansatz, we start with the basic front $u_*(x)=\tanh(x/\sqrt{2})$, which solves \eqref{e:n} at $\mu=\varepsilon=0$. We account for translations and corrections with the ansatz
\[
u(t,x)=u_*(x-\xi(t))+w(t,x-\xi(t)),
\]
which yields the equation
\[
-\dot{\xi}u_*'+w_t-w_y\dot{\xi}=w_{yy}+f'(u_*)w +h(y,w;\mu)+\varepsilon\alpha(y+\xi;\theta)g(u_*+w),
\]
where $f(u)=u(1-u^2)$ and 
\[
h(y,w;\mu)=f(u_*+w)-f(u_*)-f'(u_*)w+\mu(1+ u_*+w)(1-u_*-w)=\rmO(|w|^2+|\mu|).
\]
The linear operator $L=\partial_{yy}+f'(u_*)$ is self-adjoint with kernel $u_*'$, and we will normalize $w$ through
\[
\langle w,u_*'\rangle:=\int_\R w(t,y)\cdot u_*'(y)\rmd y=0,
\]
which implies $\langle w_t,u_*'\rangle=0$. Decomposing with the orthogonal projection onto $u_*'$, we obtain
\begin{align*}
\dot{\xi}&=  -\frac{1}{\langle u_*'+w_y,u_*'\rangle}
\left\langle u_*',h(y,w;\mu)+\varepsilon\alpha(y+\xi;\theta)g(u_*+w)\right\rangle   
\\
w_t&=Lw+h(y,w;\mu)+\varepsilon\alpha(y+\xi;\theta)g(u_*+w)+\dot{\xi} (u_*'+w_y),
\end{align*}
where in the equation for $w_t$, we substitute the expression from the first equation for $\dot{\xi}$ to obtain a system of evolution equations for $(\xi,w)$.

Clearly, given $\xi$ and $w$, we can reconstruct $u$ and vice-versa. Translation symmetries act trivially in these new coordinates, $T_\zeta(\xi,w)=(\xi+\zeta,w)$. Since $\alpha(y+\xi;\theta)=\alpha(y;S_{\xi}(\theta))$, the right-hand side of the $w$-equation depends on $\xi$ only through its dependence on $S_{\xi}$. 

We pose this equation in a space of exponentially localized functions $w(t,\cdot)\in BC^0_{-\delta}$, where $\delta$ is sufficiently small such that $u_*'\in BC^0_{-\delta}$. Since $g$ and $g'$ vanish at $\pm1$, the map $w\mapsto g(u_*+w)$ is 
smooth from $BC^0_{-\delta}$ to $BC^0_{-2\delta}$. Multiplication with $\alpha$ yields a smooth map $(\theta,w)\in \mathcal{M}\times BC^0_{-\delta}\to BC^0_{\eta-2\delta}$. For tail-Lyapunov dominance, $\delta\geq\eta$, the nonlinearity therefore defines a smooth automorphism on $BC^0_{-\delta}$ with $C^2$-dependence on $\theta$, by Hypothesis \ref{h:l0}.

A standard contraction mapping theorem \cite{henry}, now gives the existence of a center manifold smoothly depending on the parameters $\mu$ and $\theta$ with induced flow respecting the symmetry. Expansions for the reduced vector field follow by projecting the leading order terms in $\varepsilon$ and $\mu$ onto the eigenspace, which here is equivalent to computing the leading order terms of the $\xi$-vector field. 
\end{Proof}
\begin{Remark}[vanishing tail vs Lyapunov growth] It is clear from the proof that tail-Lyapunov dominance can be weakened when assuming higher order of vanishing tail corrections, say, $g^{(j)}(\pm 1)=0$, $0\leq j\leq \ell$, which readily gives a required tail-Lyapunov dominance relation of $\eta\leq\ell \delta$. 
\end{Remark}
\begin{Remark}[quasiperiodic media] When the flow $S$ is simply irrational rotation on a torus, one readily sees that trajectories are bounded  and Hypothesis \ref{h:11}  on vanishing tail corrections is not needed.
\end{Remark}
One can clearly construct examples of nonlinearities satisfying Hypothesis \ref{h:11} quite easily, using for instance $g(u)=(u^2-1)^2$. A simplest example for Hypothesis \ref{h:1} are quasiperiodic flows on $\T^\kappa$ with zero Lyapunov exponent $\eta$ and invariant ergodic Lebesgue measure, for instance 
\begin{equation}\label{e:irr}
\alpha(y;\theta)=\sum_{j=1}^\kappa \alpha_j\cos(\omega_j y+2\pi \theta_j),
\end{equation}
with $(\omega_j)_{j=1,\ldots,\kappa}$
independent over the rationals, and $\alpha_j\neq 0$.

\begin{Remark}[ergodic media beyond quasiperioidicity]
 More intriguing examples can be constructed from suspensions of ergodic diffeomorphisms. Consider a diffeomorphism on a compact manifold $\Psi:\mathcal{M}_0$ and define $\mathcal{M}:=(\mathcal{M}_0\times [0,1])/\sim $, where the equivalence relation identifies $(1,\theta)$ and $(0,\Psi(\theta))$. The suspension flow $S_\zeta$ simply translates along the interval $[0,1]$, such that $S_1(\theta,0)=(\Psi(\theta),0)$. An ergodic measure $\nu_0$ for $\Psi$ induces an ergodic product measure for the suspension flow $\nu=\nu\times \nu_\R$, where $\nu_\R$ is Lebesgue measure on $[0,1]$, augmenting the dimension of $\nu_0$ by 1. Anosov diffeomorphisms such as the cat map $(u,v)\mapsto (2u+v,u+v)$ on $\R^2/\Z^2$ are examples of ergodic maps with in this particular case, 2-dimensional ergodic Lebesgue measure. Lower-dimensional examples can be obtained from horseshoe maps; see for instance \cite{young}. For the simplest horseshoe and precisely (affine) linear dynamics on the invariant set, one finds 
 \[
 \kappa_0=\mathrm{dim}\,\nu_0=(\log 2 )\left(\frac{1}{\log(\rho^\mathrm{u})}-\frac{1}{\log(\rho^\mathrm{s})}\right),
 \]
 where $\log 2 $ is the entropy (more generally given through $\log$($\#\{$crossings$\}$)), and $\rho^\mathrm{u/s}$ are expansion and contraction rates, respectively. To see this, one first notices that the ergodic measure of maximal dimension is a product measure. One then uses invariance of the measure to see that the measure of a narrow vertical stripe equals the measure of two vertical stripes with width contracted by $\rho^\mathrm{s}$, which gives, after iteration, $\nu(S_r)=(1/2)^n$ with $r=(\rho^\mathrm{s})^n$ for vertical strips $S_r$ of width $r$. A similar consideration for vertical strips and backward iteration then yields the desired result for the dimension.
 
 Noting that $\rho^\mathrm{u}>2$ and $\rho^\mathrm{s}<1/2$, we note that we can realize arbitrary dimensions $\kappa_0\in (0,2)$ and hence ergodic dimensions $\kappa\in (1,3)$ for the suspension flow. 
 
 Abandoning invertibility of $S_\xi$ and focusing on $t\to +\infty$, simpler examples can be constructed from suspensions of (expanding) interval maps such as $x\mapsto 2x\mod 1$ or $x\mapsto 4x(1-x)$.  
\end{Remark}
\begin{Remark}[hypothesis on depinning]
Hypothesis \ref{h:2} can be realized under suitable assumptions on the inhomogeneity $\alpha(y;\theta)$.With the example from irrational media \eqref{e:irr}, $g(u)=(u^2-1)^2$,  one finds 
\[
s_\varepsilon(\theta)=\int_\R \frac{3}{4}\mathrm{sech}\,(x/\sqrt{2})^6 \sum_{j=1}^\kappa \alpha_j\cos(\omega_j y+2\pi \theta_j)\rmd y=\sum_{j=1}^\kappa\alpha_j\frac{\pi\omega_j(2+\omega_j^2)(8+\omega_j^2)}{20} \mathrm{csch}\,(\pi\omega_j/\sqrt{2})\cos(2\pi\theta_j),
\]
which is an expression of the form $\sum_j \beta_j\cos(2\pi\theta_j)$ with nondegenerate minimum  $\theta=0\in \T^\kappa$ for $\beta_j<0$, say.
\end{Remark}
%
%
%
%

\section{Depinning  --- numerical corroborations}\label{s:4}

We expect the results from the abstract framework in Section \ref{s:2} to be applicable in a much wider context than guaranteed by the results in Section \ref{s:3}. We therefore tested the predicted asymptotics in the context of  lattice-dynamical systems. Due to the discrete nature of translation symmetry, motion of traveling waves here is inherently periodic, such that the dimension of the medium needs to be increased by one. Considering for instance a lattice-differential equation 
\begin{equation}\label{e:lds2}
\dot{u}_j=d(u_{j+1}-2 u_j+u_{j-1})+(u_j-a_j)(1-u_j^2),
\end{equation}
one would consider the dimension of the set of translates of the sequences $(a_j)_{j\in\Z}$ in a local topology, and add one, to obtain our dimension $\kappa$. The additional dimension can also be understood from the results in \cite{vVs}, where lattice dynamical systems were approximately embedded into reaction-diffusion equations with spatially periodic coefficients. In that respect, a constant, translational invariant lattice dynamical system $a_j\equiv a$, would correspond to a periodic medium, $\kappa=1$. 

\begin{figure}[h!]
\includegraphics[width=0.32\textwidth]{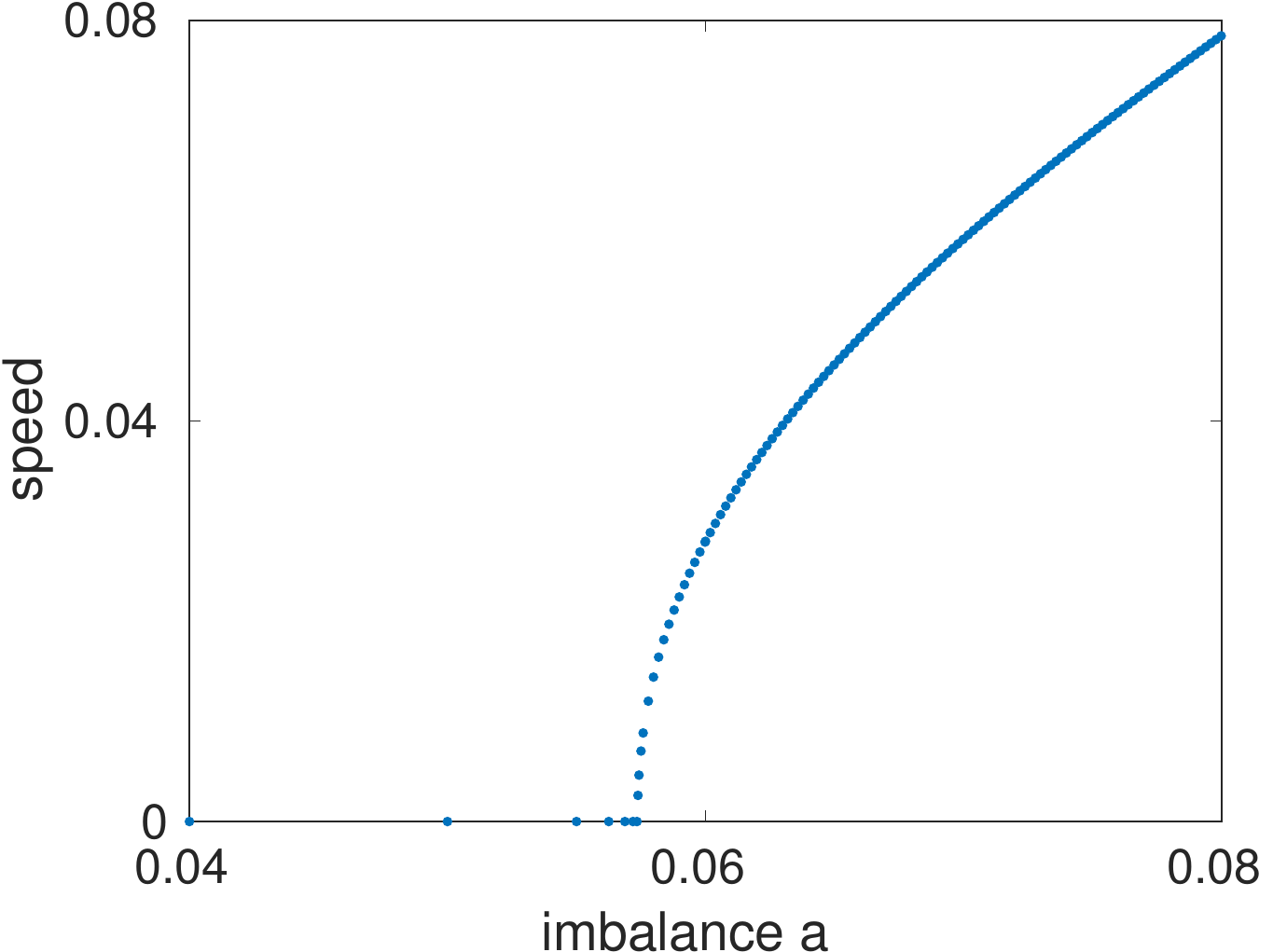}\hfill
\includegraphics[width=0.32\textwidth]{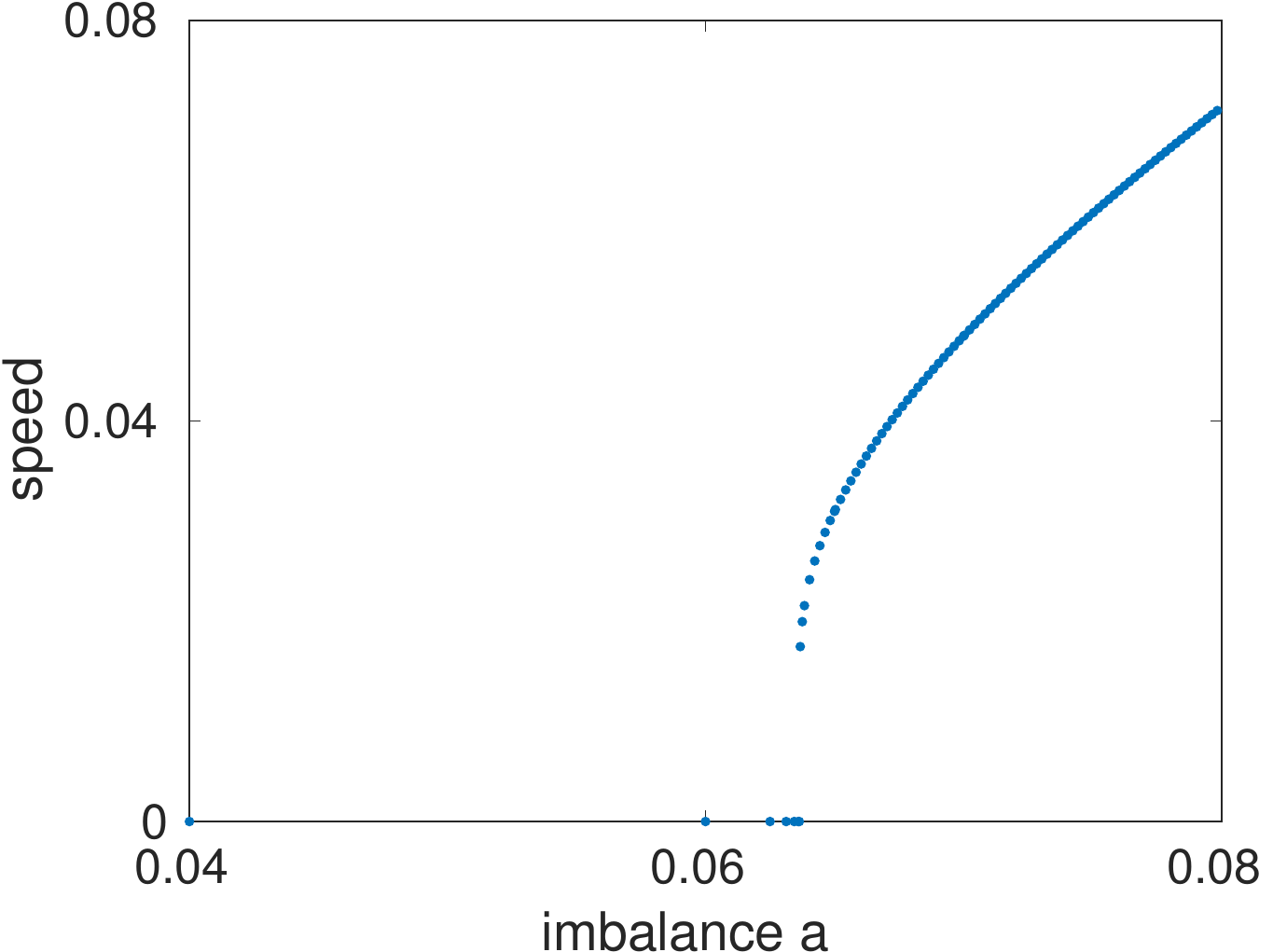}\hfill
\includegraphics[width=0.304\textwidth]{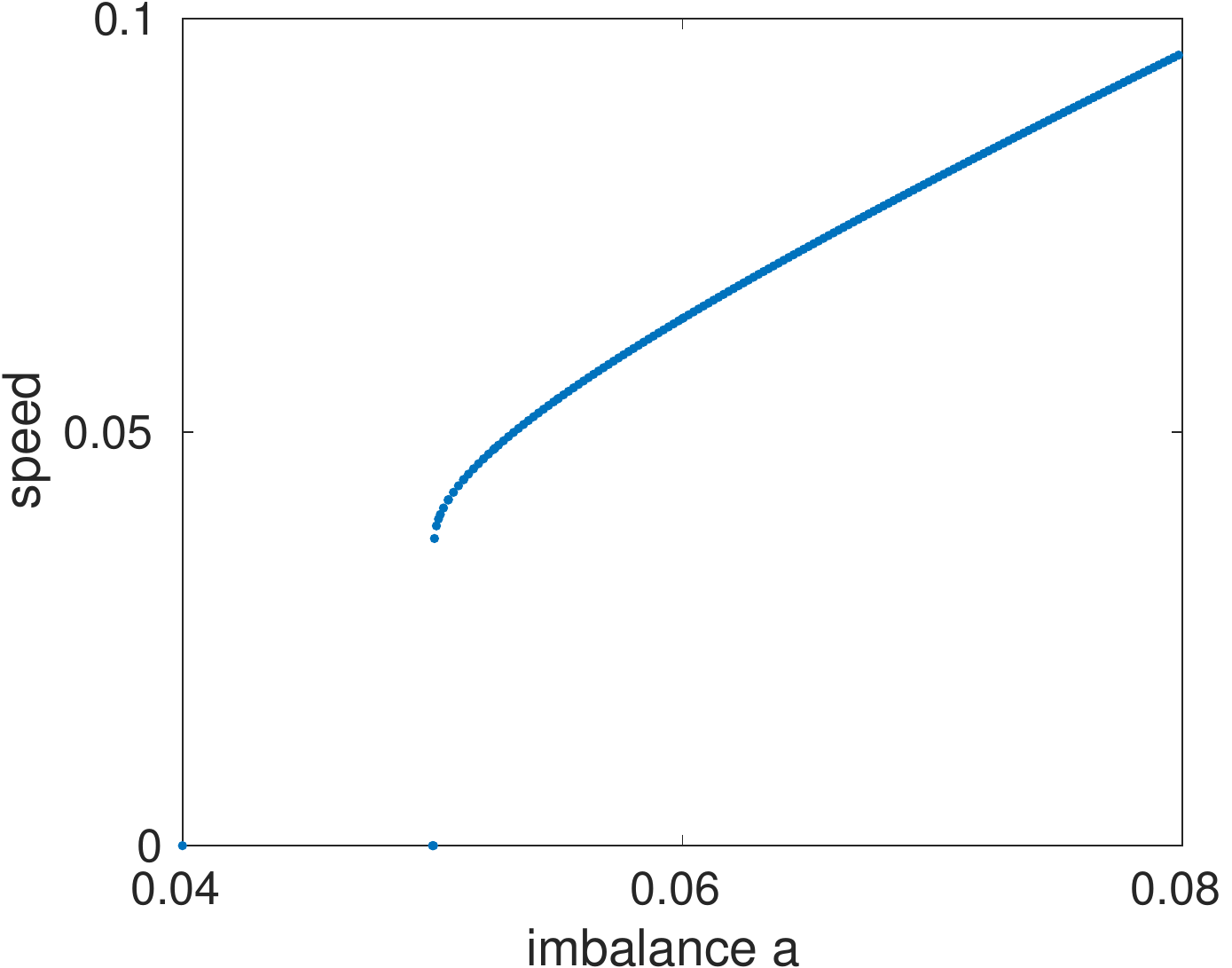}\\[0.1in]
\centering\includegraphics[width=0.306\textwidth]{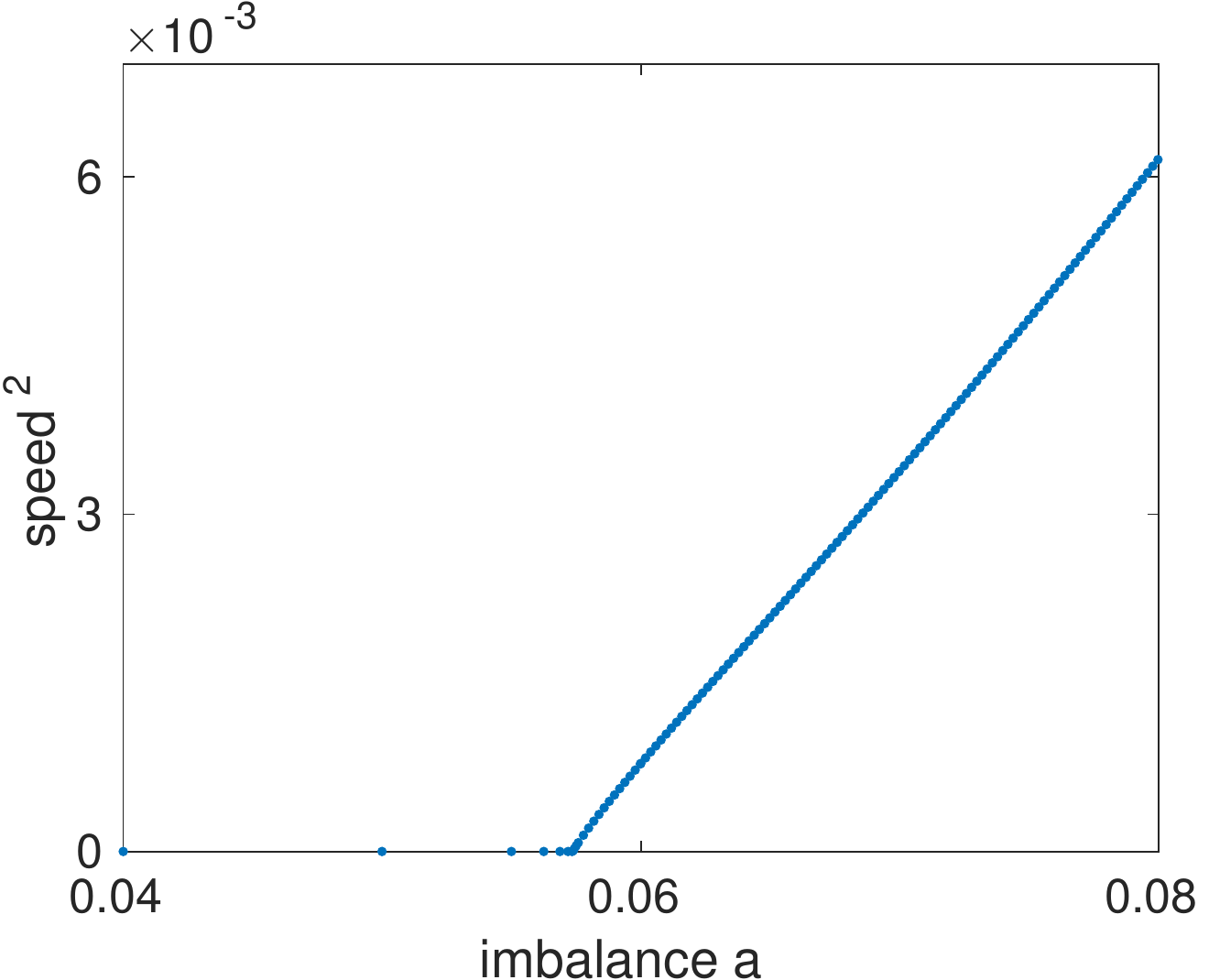}\hspace*{0.1in}
\includegraphics[width=0.32\textwidth]{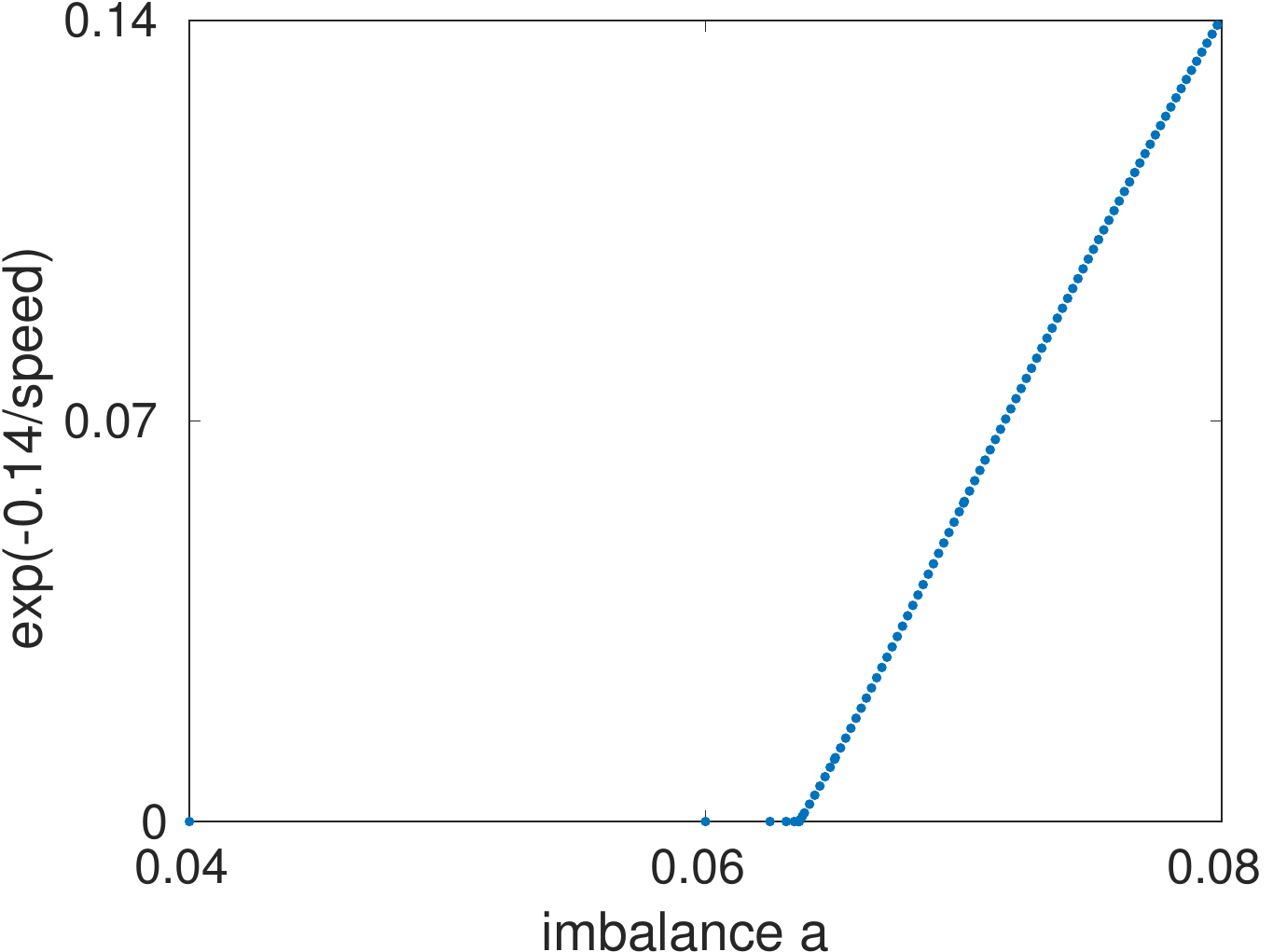}
\caption{Speed-imbalance relations computed for the lattice-dynamical system \eqref{e:lds}, with spatially periodic, 2-frequency, and 3-frequency media from left to right, top row, as specified in the text; plot of  $\bar{s}^2$ for periodic and $\exp(-c/\bar{s})$ with visual best fit $c=0.14$ exhibit linear asymptotics as predicted by our analysis. }\label{f:4}
\end{figure}

We performed numerical experiments for \eqref{e:lds2} 
with various choices for the $a_j$. Since we expect results to be robust across many systems, we used rough numerical discretization, explicit Euler with step size $h=0.1$, on a finite-dimensional approximation with $N=71$ points. We used appropriate shifts to keep the moving interface in the center of the domain; see \cite{cb} for an equivalent setup. We ran simulations for times that amount to about $500$ effective shifts on the lattice.

Specifically, we chose
\begin{description}
  \item \emph{Periodic:} $a_j = a + \varepsilon \cos(\omega j + \theta)$, with $\varepsilon = 0.1$, $\omega = \pi/2$, $\theta = 0$;
  \item \emph{2-Frequency:}   $a_j = a + \varepsilon \cos(\omega j + \theta)$, with $\varepsilon = 0.1$, $\omega = \sqrt{2}$, $\theta = \sqrt{3}$
  \item \emph{3-Frequency:} $a_j = a + \varepsilon \cos(\omega_1 j + \theta_1)+ \varepsilon \cos(\omega_2 j + \theta_2)$, with $\varepsilon = 0.05$, $\omega_1 = \sqrt{2}$, $\theta_1 = \sqrt{3}$, $\omega_2 = \sqrt{5}$, $\theta_2 = \sqrt{6}$.
\end{description}
The results are shown in Figure \ref{f:4}. The plots show the steepening of speed-imabalance relations as the dimension of the medium increases. We also plotted $\bar{s}^{1/\kappa}$ in the periodic medium, which according to our prediction should exhibit linear asymptotics. For the critical dimension $\kappa=2$, 2-frequency medium, we plotted $\exp(-c/\bar{s})$, which should again exhibit linear asymptotics when $\bar{s}\sim c|\log\mu|$. We found $c=0.14$ for a best visual fit. More rigorously fitting the parameter $c$ would require a more accurate computation of the depinning threshold $a_\mathrm{c}$.

\section{Summary and extensions}\label{s:5}
 
After summarizing our point of view, we comment on a number of open questions related to the results here. 

\paragraph{Summary of results.}
Our goal was to derive universal scaling laws near depinning transitions. Our main result distills a framework in which depinning asymptotics are governed by power laws with exponent depending on the local dimension of the invariant measure near criticality. The framework is inspired by symmetry considerations, viewing translations of the front as equivalent to shifts of the medium. Those shifts of the medium are naturally viewed as a flow, which, as our main assumption, we view as being induced by a smooth flow on a smooth compact manifold, with an ergodic measure capturing the statistics of translates of the medium when considered in a local topology. The emphasis on symmetry and reduction to skew-product flows is inspired by \cite{fssw,ssw}, where motion on groups was forced by ``internal'' dynamics of fronts or other coherent structures. 
As such, our results rely crucially on smoothness of extensions. We view ergodicity as somewhat more natural, given that flows on a manifold always possess ergodic measures. 
When constructing the reduced skew-product flow in a more concrete example, we noticed a subtle condition, requiring bounds on spatial Lyapunov exponents of the medium in terms of exponential convergence rates of the front.

\paragraph{Critical fronts.} At $\mu=0$, the critical threshold for depinning, the medium supports a pinned front --- in the support of the ergodic measure, therefore not necessarily in the given fixed medium. The rate of growth of $\xi$ depends very much on the dimension and on the particular medium. Clearly, trajectories are bounded in one-dimensional media. In quasi-periodic media, the orbit is bounded whenever the critical medium $\theta_*$ is a specific shift of the given medium, that is, the maximum lies on the given trajectory on the torus. For most media $\theta$, however, this will not be the case, and the front position will not be bounded. For dimensions $\kappa>2$, the propagation will be ballistic $\xi\sim t$,  with limiting speed the continuous limit of speeds for $\mu>0$, as one can readily see from the calculation of the speed, exploiting that the singular integral converges for $\mu=0$. For dimensions $\kappa\leq 2$, we expect propagation to be sub-ballistic. Heuristically, times spent near the pinned front scale with $1/|\psi|$, where $\psi$ is a variable parameterizing a section to the flow for $\psi$ on $\mathcal{M}$ near the pinned value $\theta_*$; see Remark \ref{r:geo} for a description of geometry. We then expect that $\xi=N$ at times $T\sim \sum_{j=1}^N \psi_j^{-1}$, which, for $\psi_j$ asymptotically uniformly distributed according to a $\kappa-1$-dimensional measure in the section, gives 
\[
T\sim N\int_{r=N^\frac{-1}{\kappa-1}}^1\frac{1}{r^{\kappa-2}}\rmd r\sim N\cdot N^\frac{2-\kappa}{\kappa-1} \mbox{ for } \kappa\in(1,2),
\]
which then gives
\begin{equation}\label{e:crit}
\xi\sim T^{\kappa-1}, \qquad 1<\kappa<2.
\end{equation}
A similar calculation for $\kappa=2$ gives 
\begin{equation}\label{e:ccrit}
\xi\sim T/|\log(T)|.
\end{equation}

\paragraph{Sharp asymptotics.} In the case of two-dimensional quasi-periodic media, the expansion coefficient is in fact explicit, given quadratic terms of the minimum and the dependence on $\mu$. Consider therefore the inhomogeneity $(1-u^2)^2\sum_{j=1}^2\alpha_j\cos(\omega_j x + \theta_j)$, which gives the reduced vector field expansion
\[
{s}_\varepsilon(\theta)=\sum_{j=1}^2 \beta_j \cos(2\pi\theta_j),\qquad {s}_\mu=\sqrt{2},
\]
with 
\[
\beta_j=\alpha_j \frac{\omega_j(2+\omega_j^2)(8+\omega_j^2)\pi}{20}\mathrm{csch}\,(\pi\omega_j/\sqrt{2}).
\]
Assuming $\alpha_j<0$, we find minima at $\theta_j=0$, which leads to depinning thresholds
$
\mu_\mathrm{c}=\varepsilon \sum\beta_j
$ and to an expansion $s(\theta)=\sqrt{2}\mu -\sum2\pi^2\beta_j\theta_j^2$, with resulting depinning asymptotics for the ergodic integral 
\[
\bar{s}(\mu)= 16\pi\sqrt{\beta_1\beta_2}|\log(\mu)|^{-1}.
\]
Similar calculations are possible whenever the measure $\nu$ is known explicitly.

\paragraph{Higher asymptotics.} Beyond ballistic asymptotics, corresponding to the ergodic average, one could ask for rates of convergence. In quasi-periodic media, one encounters subtle dependence on frequencies \cite[\S 2,3]{kn}, with convergence $\xi(T)/T\sim \bar{s}+\rmO(\log T/T)$ for ``good'' irrational numbers, hence not quite asymptotic phase to an appropriately shifted uniformly translating front of speed $\bar{s}$. The correction to $\xi(T)$ is usually referred to as the discrepancy, for which $\log T/T$ bounds are optimal, and bounds $T^\delta/T$ are common for (positive measure) diophantine numbers.

%
%

For ``chaotic'' media, we expect Brownian deviations, as is common for ergodic averages; see  \cite{ce} for general background and  \cite{nolen} for results on fronts. Deviations can, however, be arbitrarily slowly decaying \cite{krengel}.

\paragraph{Extensions.} There clearly is a multitude of possible extensions. In immediate generalizations, one could focus  on averages as $t\to +\infty$, only, allowing for ``heteroclinic'' media with ergodic measures on the limit sets with respect to right shifts in a local topology. One could also allow more directly for random samplings of the medium. Generalizations that affect the scalings more directly are  degenerate minima, or, more interestingly, minima that occur at the boundary of the support of $\nu$, such that $\nabla s\neq 0$ at $\theta_*$. In that case, one could obtain integrals of the type $\int_0^1 r^{\kappa-1} \frac{1}{\mu+r}\rmd r$, resulting in asymptotics $\mu^{1-\kappa}$. 

Beyond the more narrow scope of one-dimensional inhomogeneous media, one could look at time-dependent media, periodic, quasi-periodic, or random, or even consider higher dimensional wave-fronts;  see for instance \cite{hoff} for results in lattices. In a different direction, motion of localized pulses in higher-dimensional media presents intriguing other possibilities, such as determining the direction of drift. In this context,  relaxation towards translational Goldstone modes may be much slower, due to interaction with continuous spectra, present whenever the ``traveling wave'' is not spatially localized. In this direction, one may also wish to study the effect of long-range interactions, such as through nonlocal coupling $K*u$, $K$ a weakly localized convolution kernel. 

More modestly, one would also wish to establish the validity of the hypotheses used in our main theorem beyond a weak inhomogeneity context, or, at least, specify numerical computations that would allow for a (semi-)rigorous verification.

%
%
%
%
%
%

\end{document}